\documentclass[12pt,a4paper]{article}
\usepackage{amsmath}
\newcommand{\rl}{{\cal F}}
\newcommand{\bx}{{\bf x}}
\begin{document}
\renewcommand{\theequation}{\arabic{section}.\arabic{equation}}
\begin{titlepage}
\title{Functionals linear in curvature \\ and statistics of helical proteins}
\author{A. Feoli\thanks{E-mail: feoli@unisannio.it; fax:
+39089965275}
\\{\it \normalsize Dipartimento di Ingegneria, Universit\`{a} del Sannio,}\\
{\it \normalsize Corso Garibaldi n.\ 107, Palazzo Bosco
Lucarelli,}\\{\it  \normalsize  82100 Benevento,  Italy}\\{\it
\normalsize and INFN Sezione di Napoli, Gruppo collegato di
Salerno }\\{\it  \normalsize  80126 Napoli, Italy} \and
 V.V. Nesterenko\thanks{E-mail: nestr@thsun1.jinr.ru} \\
{\it  \normalsize  Bogoliubov Laboratory of Theoretical Physics}
 \\{\it  \normalsize  Joint Institute for Nuclear Research}\\
{\it \normalsize   141980 Dubna, Russia}
\and G. Scarpetta\thanks{E-mail: scarpetta@sa.infn.it} \\
{\it   \normalsize     Dipartimento di Fisica "E.R.Caianiello" --
Universit\`a di Salerno}
 \\ {\it  \normalsize  84081 Baronissi (SA), Italy}
 \\{\it
\normalsize and INFN Sezione di Napoli, Gruppo collegato di
Salerno }\\{\it  \normalsize  80126 Napoli, Italy}}
\date{\empty}
\maketitle

\begin{abstract}
The effective free energy of globular protein chain is considered to be a
functional defined on smooth curves in three dimensional Euclidean
space. From the requirement of geometrical invariance, together
with basic facts  on conformation of helical proteins and
dynamical characteristics of the protein chains, we are able to
determine, in a unique way, the exact form of the free energy
functional. Namely, the free energy density should be  a linear
function of the curvature of curves on which the free energy
functional is defined. We briefly discuss the possibility   of
using the model proposed in Monte Carlo simulations of exhaustive
searching the native stable state of the protein chain. The
relation of the model proposed to the rigid relativistic particles
and strings is also  considered.
\end{abstract}


\end{titlepage}
 \section{Introduction}
A fascinating and open question challenging physics, biochemistry
and even geometry is the presence of highly regular motifs such as
$\alpha-helices$ and $\beta-sheets$ in the folded state of
biopolymers and proteins. A wide range of approaches have been
proposed to rationalize the existence of such secondary structures
(see, for example, reviews \cite{CD,Dill,BMMS} and references
therein).

 In principle, in order to find the
stable native state of a protein, one should compute, for every
possible conformation of the chain, the sum of the free energies of
the atomic interactions within the protein as well as with the
solvent and then find the conformation with the lowest free energy.
However, it is not feasible, because the number of conformations of a
protein chain grows exponentially with the chain length.

     In the present paper we are going to propose a pure geometrical
approach to describe the effective free energy of proteins, proceeding from the
most general invariance requirements and basic experimental facts concerning
the protein conformation.  It should be noted here that we shall consider not
the Hamiltonian $H$ (or energy) of an individual protein molecule, but the
effective free energy of an assembly of noninteracting protein chains. We
define this energy as such a functional $F$ depending on the radius vector
${\bf x}(s)$ of a protein chain, the stationary values of which (the extrema)
are provided only by the observed configurations of proteins. Obviously, the
effective free energy defined in such a way is analogous to the effective
action in quantum field theory~\cite{Ramon}.

Let  $E[{\bf x}(s)]$ be the energy of an individual protein chain then
the effective free energy $F$ we are interested in is apparently
determined by the functional integral
\begin{equation}
\label{fi}
e^{-\beta F[\bx_{cl}]}=\int D\bx_{qu}e^{-\beta E[\bx_{cl}+\bx_{qu}]}{,}
\end{equation}
where $\beta$ is inverse temperature and we have split the protein
radius vector $\bx$ into the classical part $\bx_{cl}$ and the
quantum part $\bx_{qu}$.  In order to use this formula the energy
functional $E$ for an individual protein chain should be specified.
The problem under consideration is close to the statistics of
semi-flexible polymers. In the last case a number of models (i.e.,
particular functionals $E$) have been proposed which substantially
rely on the differential geometry (Kratky-Porod model~\cite{K-P},
Dirac model~\cite{Dirac,Dirac-1}, tube model~\cite{tube} and so on).
Differential geometry provides a direct relation
between  physics and mathematics and thereby opens a straight route
for applying the mathematical analysis to this
physical problem~\cite{Kamien,Hyde}.
If $k(s)$ and $\kappa(s)$ are respectively the curvature and the
torsion of a polymer chain then the energy $E$ can be represented as
a polynomial in these geometrical invariants, for example, in the
form~\cite{Kamien}
\begin{equation}
\label{energy}
E=\int_{0}^{L}ds\left [
\frac{A}{2}k^2+\frac{C}{4}k^4-\alpha k^2\kappa
+ \frac{\beta}{2}k^2\kappa^2
\right ]{,}
\end{equation}
where $A,\;C,\;\alpha$, and $\beta$ are phenomenological parameters.

    However evaluation of the functional integral in Eq.\ (\ref{fi})
with the energy $E$ given in Eq.\ (\ref{energy}) or by analogous
formula is a nontrivial mathematical problem. The progress on this
line is possible only when applying approximate methods~\cite{Kamien}.
As a rule, the results obtained here do not allow one to predict the
exact stationary form  of protein or polymer chains, at best one can
calculate the statistical momenta of the chains.

In this situation we propose another way of consideration. We shall
try straight away to determine  the effective free energy
$F[\bx_{cl}]$ without specifying the  energy functional $E$ of an
individual protein chain. For this purpose we shall use experimental
knowledge concerning the protein chain configurations and the tools
of differential geometry.

   Such a statement of the problem  on describing the statistical
properties of  proteins semi-flexible polymers  is non-standard one.
In some sense it is an inverse with respect to a traditional
formulation of the problem.  Nevertheless we suppose that such an
inverse task is undoubtedly  of substantial interest.  It turns out
that when confining oneself to helical proteins the inverse problem
at hand has a nice mathematical solution, being unique. It
is the prove of this assertion that is the basic result of this paper.

 Taking into account the one-dimensional nature of the
protein chains, the relevant macroscopic free energy $F$ should be
considered as a functional defined on smooth curves ${\bf x} (s)$ (or
paths) in the three dimensional Euclidean space
\begin{equation}
F = \int {\cal F}
[{\bf x} (s)] d s,
\end{equation}
where $s$ is the length of a protein molecule. For shortennig the formulas we
drop the  subscript {\it cl}  in $\bx $.
The reparametrization invariance of the functional $F$ demands the
free energy density $\cal F$ to be  a scalar function depending on
the geometrical invariants of the position vector ${\bf x} (s)$,
which describes the spatial shape of the protein chain. In three
dimensional ambient space a smooth curve has two local invariants:
curvature $k(s)$ and torsion $\kappa(s)$. In the general case of $D$
dimensional Euclidean embedding space there are $D - 1$ principal
curvatures $k_\alpha (s), \alpha = 1,2....,D-1$ of a curve, where
$k_1 (s) = k(s)$ and $k_2(s) = \kappa(s)$.

The basic theorem (due to O.~Bonnet) in the theory of curves
reads~\cite{Geom}: given the curvatures $k_\alpha (s), \alpha = 1,2....,D-1$,  a
curve ${\bf x} (s)$ is determined in $D$-dimensional Euclidean
space up to its rotations and displacements as a whole. The
curvature $k_\alpha (s)$ is expressed in terms of the derivatives,
with respect to $s$, of the radius vector of a curve ${\bf x} (s)$
till the $(\alpha + 1)$-th order included.

The first principal curvature, or simply the curvature,  $k_1 (s) =
k(s)$ of a curve characterizes the local bending of the curve at the
point $s$. Hence, the dependence of free energy density $\cal F$ on
$k(s)$ specifies the resistance of a protein chain to be bent. The
second curvature or torsion $\kappa(s)$ is determined by the relative
rotation, around the tangent $d{\bf x}(s)/ds$ at the point $s$, of
two neighbor infinitely short elements of the protein chain. It is
well known~\cite{CD} that, in the case of protein molecules, such a
rotation is quite easy, as it requires little effort. In other words,
this rotation results in small energy differences, allowing many
overall conformations of a protein chain to arise. Thus the
dependence of the free energy density $\cal F$ on torsion $\kappa(s)$
can be neglected at least as a first approximation. Some comments
concerning this restriction will be given in Section 4. Finally one
can consider the free energy density $\cal F$ to be a function only
of the curvature $k(s)$, i.e.\ $ {\cal F} = {\cal F} (k(s))$. In what
follows we shall try to specify this dependence explicitly keeping in
mind the description of globular protein conformation.

A peculiarity of conformation of globular proteins is that they can
be ordered assemblies either of helices or of sheets as well as a
mixture of helices and sheets~\cite{CD}. In the phenomenological
macroscopic approach, which is developed here, the presence of sheets
in the spatial structure of globular proteins implies the necessity
to introduce, in addition to space curves ${\bf x} (s)$, new
dynamical variables ${\bf y} (s, s')$ describing surfaces in ambient
space.\footnote{ We do not consider here such exotic curves as
Peano's curve which, being specified by one parameter $s$,
nevertheless covers, completely tightly, a finite region of a surface
(see, for example, Ref.~\cite{SZ}).} Obviously such an extension of the
problem setting would complicate considerably our consideration.
Therefore we confine ourselves to helical proteins and try to answer
the question: Is it possible to specify the function $ {\cal F}
(k(s))$ in such a way that the extremals of the functional $ F = \int
{\cal F} d s $
 would be
only helices?  The answer to this question turns out to be
positive and unique, namely, the density of the free energy  $
{\cal F} (k(s))$ should be a linear function of the curvature
$k(s)$:
\begin{equation}
\label{result}
\rl (k)=\alpha-|\beta|\,k(s),
\end{equation}
where the constants $\alpha$ and $\beta$ are the
parameters specifying the phenomenological model proposed, $\alpha
>0,\; \beta<0$. A rigorous proof of this assertion is the main result
of the present paper.

The layout of the paper is as follows. In Section 2 the
Euler-Lagrange variational equations for the functional $F$ are
reformulated in terms of the principal curvatures. In Section 3
the explicit integrability of these equations will be shown. In
other words, for an arbitrary function $ {\cal F} (k(s))$, the
corresponding Euler-Lagrange equations  are integrable by
quadratures. The physical meaning of the relevant constants of
integrations is also discussed. In Section 4 the requirement that
the solutions of the Euler-Lagrange equations describe only
helices immediately confines the admissible free energy density  $
{\cal F} (k(s))$ to be a linear function (\ref{result}). In conclusion
(Section 5), the obtained results and their physical implications
are briefly discussed.

 \section{Euler-Lagrange equations in terms of principal curvatures}
\setcounter{equation}0

Keeping in mind the description of the protein conformation, one
should consider the problem in three dimensional Euclidean space.
However, from the mathematical standpoint, it is convenient to
formulate the equations we are looking for, first in
$D$-dimensional Euclidean space and  only after that to go to the
three dimensional ambient space.

We shall follow our papers~\cite{NFS1,NFS2}, where the analogous problem has
been considered in Minkowski space-time and the results obtained
have been applied to the models of relativistic particles
described by the Lagrangians dependent on higher derivatives of
the particle coordinates. Unfortunately, the transition to the
Euclidean space in the final formulas of Ref.~\cite{NFS1} is not
obvious.\footnote{Differential geometries of curves in Euclidean
space and in Minkowski space-time differ substantially. In main
part it is due to three types of curves existing in
pseudo-Euclidean space: timelike, spacelike and isotropic (null)
curves~\cite{NRG}.} Therefore we derive the equations for the principal
curvatures and the relevant integrals of motion in Euclidean space
anew. In this section we shall use a natural parametrization of
the curve $\bf x (s)$ in the $D$-dimensional Euclidean space:
$x^i(s),  \;  i = 1,2,  \ldots  ,  D$. In this parametrization
\begin{equation}
     \frac{dx^i}{ds}\,\frac{d x^i}{ds} = ({\bf x}'\, {\bf x}') = 1,
     \end{equation}
or in another way $ds  ^2 = dx^i dx^i = (d{\bf x}\, d{\bf x})$. As
usual the sum over the repeated indexes is assumed in the
corresponding range and, for shortening, the differentiation with
   respect to the natural parameter $s$ will be denoted by a prime.

The curve  $x^i (s),  \;  i  = 1,2,  \ldots  ,  D$ has $D-1$
principal (or external) curvatures which determine this curve up
to its motion as a whole in embedding space. The first curvature,
or simply curvature, is given by
\begin{equation}
     k_1^2(s) = k^2 (s) =
      \frac{d^2 x^i}{ds^2} \frac{d^2 x^i}{ds^2} = ({\bf x}'' {\bf x}'') {.}
     \end{equation}

For an arbitrary functional $F$ defined on  curves $x^i (s)$ in
$D$-dimensional space the Euler-Lagrange equations are a set of
exactly $D$ equations
\begin{equation} \frac{\delta F}{\delta x^i} = 0, \quad i= 1,2,\ldots ,D.
 \end{equation}

 However, if the functional $F$ depends
only on the curvature
\begin{equation} F = \int {\cal F} (k(s)) ds, \end{equation}
then $D$ equations (2.3) for $D$ variables $x^i (s),  \;  i   =
1,2, \ldots  ,  D$ give $D-1$ equations for the principal
curvatures $k_\alpha (s),  \;  \alpha   =  1,2,  \ldots  , D-1.$

When deriving the equations for $k_\alpha (s)$ we shall use the
orthonormal Frenet basis $\{ {\bf e}_a\}$:
\begin{equation}
 ({\bf e}_a \,{\bf e}_b) = \delta_{a b}, \quad  a, b = 1,2....., D
\end{equation}
associated with the curve ${\bf x} (s)$ and the Frenet equations
governing the motion of this basis along the curve ${\bf x} (s)$
\begin{equation}
\frac{d{\bf e}_a}{ds}  = \omega _{ab} \,{\bf e} _b, \quad \omega
_{a b}+\omega _{b a} = 0\,{.}
\end{equation}
The unit vector of the Frenet basis ${\bf e}_\alpha (s)$
 is expressed in terms of the derivatives of
the position vector ${\bf x} (s)$ with respect to the parameter
$s$ till the order $\alpha$ included, $\alpha = 1,2, ...., D$. The
first vector ${\bf e}_1$ is directed along the tangent of the
curve at the point $s$
 \begin{equation}
   {\bf e}_1 (s)  = \frac{d{\bf x}}{ds} = {\bf x}' .
    \end{equation}
In what follows, the explicit expressions, in terms of ${\bf x}
(s)$, for the rest of the unit vectors ${\bf e}_a (s)$ are not
necessary (relevant formulas can be found, for example,
in~\cite{Postnikov, Aminov}). Nonzero elements of the matrix $\omega$ in the Frenet
equations (2.6) are determined by the principal curvatures
\begin{equation}
\omega _{\alpha ,\, \alpha + 1} =  -\, \omega _{\alpha +1,\,
\alpha} = k_{\alpha +1}(s), \quad \alpha  = 1,  \ldots ,D-1\,{.}
\end{equation}
In turn, the principal curvature $k_\alpha (s)$ is expressed in
terms of the derivatives of the position vector ${\bf x} (s)$ till
the $(\alpha +1)$-th order included. The first curvature is
defined in Eq.\ (2.2). Besides, we shall use the explicit
expression, through the derivatives of ${\bf x} (s)$, only for the
second curvature or torsion $k_2(s) = \kappa(s)$.

The variation of the space form of the protein molecule
\begin{equation}
  \delta {\bf x} (s)  =  \varepsilon_a (s)\,{\bf e}_a
(s),  \quad a  = 1, \ldots, D
\end{equation}
results in the following variation of the free energy functional
$F$ defined in Eq.\ (2.4)
 \begin{equation}
\delta F  = \delta F_1 \,+\,\delta F_2,
     \end{equation}
     where
     \begin{equation}
    \delta F_1 = \int \!\!  ds \,{\cal F}'(k_1)\,\delta k_1(s),
\end{equation}
\begin{equation}
    \delta F_2 = \int \!\!{\cal
F}(k_1)\,\delta ds.
     \end{equation}
In what follows the  prime  on the free energy density  ${\cal
F}(k_1)$ will denote the differentiation with respect to its
argument $k_1$.

The variation $\delta\, ds$ is calculated in a straightforward way
\begin{equation}
\delta \,ds = \delta\,\sqrt{dx^i \,dx^i } = \frac{dx^i \,\delta\,
d x^i}{ds} = \frac{dx^i}{ds}\, d(\delta x^i ) = ({\bf x}^\prime\,
d\delta{\bf x}){.}
\end{equation}
The substitution of Eq.\ (2.13) into $\delta F_2$ and the subsequent
integration by parts\footnote{We are interested in statistical
applications of the ultimate equations, therefore the contributions
depending on the position of the protein edges are dropped.} give
\begin{equation}
\delta F_2  = -\int \!\!\rl '(k_1)\, k^{\prime}_1\,( {\bf x}^{\prime}
\delta {\bf x} )\,ds\,-
 \int \!\!
\rl (k_1)\,( {\bf x}^{\prime \prime} \delta {\bf x} )\,ds\,{.}
\end{equation}
By making use of the Frenet equations (2.6) and the expansion (2.9)
one can derive $$ ({\bf x}^{\prime} \delta {\bf x}) = ({\bf e}_1
\delta {\bf x}) = \varepsilon_a ({\bf e}_1 {\bf e}_a) =
\varepsilon_1, $$
\begin{equation}
({\bf x}^{\prime \prime}\, \delta {\bf x}) = ({\bf e}^{\prime}_1
\,\delta {\bf x}) = \omega_{12} ({\bf e}_2\, \delta {\bf x}) = k_1
\varepsilon_2.
\end{equation}
In view of this, Eq.\ (2.12) acquires the form
\begin{equation}
                   \delta F_2  = - \int \!\! \rl ' (k_1)\, k^{\prime}_1\,
 \varepsilon_1 (s)\,ds \,- \int \!\! \rl (k_1)\,k_1(s)\,
 \varepsilon_2 (s)\,ds\,{.}
\end{equation}

The calculation of the variation $\delta F_1$, defined in Eq.\
(2.11), is more complicated. First we find the variation of the
curvature $\delta k_1(s)$. Definition (2.2) and the Frenet
equations (2.6) give
$$ k_1(s)\,\delta k_1(s) = \left ({\bf x}^{\prime \prime} \delta
{\bf x}^{\prime \prime} \right)= \left( {\bf e}_{1}^{\prime} \delta
{\bf x}^{\prime \prime} \right )= k_1 ({\bf e}_{2} \delta {\bf
x}^{\prime \prime} ). $$ Hence
\begin{equation}
\delta k_1 = ({\bf e}_{2}\, \delta {\bf x}^{\prime \prime} ).
\end{equation}
Applying Eq.\ (2.13) step by step one derives
\begin{equation}
\delta{\bf x}^{\prime} = \delta\frac{d{\bf x}}{ds} =
\frac{d}{ds}\delta{\bf x} - \frac{d{\bf x}({\bf x}' d\,\delta{\bf
x})}{ds^2} = \frac{d}{ds} \delta{\bf x} - {\bf x}' ({\bf x}'
\frac{d}{ds} \delta{\bf x}) = \frac{d}{ds} \delta{\bf x} - {\bf
e}_1 ({\bf e}_1\frac{d}{ds}\delta{\bf x}),
\end{equation}
\begin{equation}
\delta{\bf x}^{\prime \prime} = \delta\frac{d{\bf x}'}{ds} =
\frac{d}{ds}\delta{\bf x}' - \frac{d{\bf x}' ({\bf x}'
d\,\delta{\bf x})}{ds^2} = \frac{d^2}{ds^2} \delta{\bf x}  - 2{\bf
e}_1^{\prime} ({\bf e}_1\frac{d}{ds}\delta{\bf x}) - {\bf
e}_1\frac{d}{ds} ({\bf e}_1\frac{d}{ds}\delta{\bf x}).
\end{equation}
In view of (2.17), the last term in Eq.\ (2.19) does not
contribute to $\delta k_1$. Substituting Eq.\ (2.19) into (2.17)
and using the Frenet equations (2.6) together with expansion (2.9)
we obtain
\begin{equation}
\delta k_1(s) = \varepsilon_1  k^{\prime}_1 + \varepsilon_2 +
\varepsilon_2 (k^2_1 - k_2^2) - 2  \varepsilon^{\prime}_3 k_2 -
\varepsilon_3 k^{\prime}_2 - \varepsilon_4 k_2 k_3{.}
\end{equation}
As one could expect formula (2.20) for $\delta k_1$ differs from
its analogue in Minkowski space (see Eq.\ (2.18) in Ref.~\cite{NFS1} only
by signs of some terms. However, without the direct calculation of
$\delta k_1$ in Euclidean space the rule of sign changes in this
expression is not obvious.

 Thus, the variation of the first curvature, $\delta k_1(s)$,
depends on the variations of the worldline coordinates only along
the directions $\;{\bf e}_1,\;{\bf e}_2, \; {\bf e}_3$, and ${\bf
e}_4$ (on the four arbitrary functions $\varepsilon_a (s),\; a
 = 1,2,3,4 )$ and on the first three curvatures $k_1,\;k_2$, and
$k_3$.

Substituting Eq.\ (2.20) into (2.11), integrating the latter
equation by parts, and taking into account (2.16) we obtain
\begin{eqnarray}
\lefteqn{ \delta F = \int \!\! ds \left \{ \left [ \left
(k_1^2-k_2^2\right )
 \rl'(k_1) +\frac{d^2}{ds^2}(\rl '(k_1))-k_1\,\rl(k_1) \right
] \varepsilon_2 (s)
 \right .} \nonumber \\
&& +\! \left .\left [ 2\,\frac{d}{ds}(\rl '(k_1)\,k_2)-
k^{\prime}_2\,\rl '(k_1)
 \right ]\!\! \varepsilon_3 (s)-\rl '(k_1)\,k_2 k_3\varepsilon_4
 (s)\!
 \right \}\! = 0{.}
\end{eqnarray}
The terms in $\delta F_1$ and $\delta F_2$, depending on the
variation $\varepsilon_1 (s)$ along the tangent to the curve, are
mutually cancelled, and the variation $\delta F$ depends, as one
could expect, only on the normal variation of the curve $ {\bf x}
(s)$, or more precisely, on the variation of $ {\bf x} (s)$ along the
three normals $\;{\bf e}_2,\;{\bf e}_3$, and $ {\bf e}_4$. Apparently
the last property of the variation $\delta F$ is due to a special
type of functionals (2.4) under consideration, which depend only on
the curvature of a curve. From Eq.\ (2.21) three equations for
principal curvatures follow
\begin{eqnarray}
\frac{d^2}{ds^2}\,(\rl'(k_1)) & =& - \left ( k_1^2-k_2^2\right
)\, \rl '(k_1)+k_1\,\rl(k_1)\,{,}\\
2\,\frac{d}{ds}\,(\rl'(k_1)\,k_2) &=&  k^{\prime}_2\,\rl'(k_1)\,{,}
\\ \rl '(k_1)\,k_2\,k_3&=&0\,{.}
\end{eqnarray}
At first glance, Eqs.\ (2.22) -- (2.24) are obviously insufficient
in order to determine all $D-1$ principal curvatures of the curve
$ {\bf x} (s)$ embedded in $D$-dimensional Euclidean space with
arbitrary $D$. However, it is not the case.

We start the analysis of the obtained equations from Eq.\ (2.24).
The most general condition on the principal curvatures, following
from this equation, is the requirement of vanishing the third
curvature
\begin{equation}
k_3 (s) = 0.
\end{equation}
The point is that the principal curvatures, due to their
construction, obey the conditions~\cite{Aminov,Grif}: if a curvature $k_j (s)$
does not vanish at the point  $s$, then at this point the curvatures
$k_\alpha (s)$ with $\alpha = 1,2,....,j-1$ are also different from
zero. But the vanishing of $k_j (s)$ at a point  $s$, $k_j (s)=0$,
implies that $k_\alpha (s) = 0$ for $\alpha = j+1, j+2,..., D-1$. In
view of this, Eq.\ (2.25) entails the following conditions
\begin{equation}
k_4 (s) = k_5 (s)= \ldots = k_{D-1} (s) = 0.
\end{equation}

 Thus, in the problem  under consideration there are two non
trivial equations (2.22) and (2.23) for the curvatures $k_1 (s)$ and
$k_2 (s)$. Equation (2.23) can be integrated with arbitrary free
energy density ${\cal F} (k_1)$. Actually, for $ k_2 \neq 0$ one can
rewrite this equation in the form $$ \frac{k^{\prime}_2}{k_2} +
2\frac{\rl '' (k_1)k^{\prime}_1}{\rl ' (k_1)} = 0 $$
 or
$$2\, d\, \left [\ln \rl '(k_1)\right ]\,+\,d\,(\ln k_2) = 0\,{.}$$
Hence
\begin{equation}
\label{2-27}
 \left ( \rl ' (k_1)\right )^2k_2=C\,{,}
\end{equation}
where $C$ is an integration constant.

 Relation (\ref{2-27}) enables one to eliminate the torsion $k_2(s)$ from Eq.\ (2.22).
 As
a result we are left with one nonlinear differential equation of
the second order for the curvature $k_1(s)$
\begin{equation}
\label{final} \frac{d^2}{ds^2}\,\rl '(k_1) + \left ( k_1^2  -
\frac{C^2}{(\rl '(k_1))^4}\right ) \,\rl '(k_1) -k_1\,\rl (k_1) =
0.
\end{equation}

Having resolved this equation for $k_1(s)$, one can determine the
rest of curvatures by making use of Eqs. (2.27), (2.25),  and
(2.26). Integration of the Frenet equations (2.6) with principal
curvatures found enables one to recover the curve $ {\bf x} (s)$
itself.

Notwithstanding its nonlinear character, Eq.\ (\ref{final}) can be
integrated in quadratures for arbitrary function ${\cal F} (k_1)$. To
show this, the first integral for this equation will be constructed
proceeding from the symmetry properties of the variational problem
under study.

 \section{Exact integrability of the Euler-Lagrange equations for
principal curvatures} \setcounter{equation}0

The functional (2.4) possesses a quite large set of symmetries,
the analysis of which will allow us to construct the first
integral for nonlinear equation (\ref{final}). In this section we
shall use an arbitrary parametrization of the curve $ {\bf x}
(\tau)$. Now the functional (2.4) assumes the form
\begin{equation}
F = \int \!\!\rl (k_1)\, \sqrt {\dot {\bf x}^2}\,d \tau\,{,}
\end{equation}
where a dot over $ {\bf x}$ denotes the differentiation with respect
to the parameter $\tau$. In arbitrary parametrization the curvature
$k_1$ is given by
\begin{equation}
k_1^2=\frac{\dot {\bf x}^2\, \ddot {\bf x}^2-(\dot {\bf x}\ddot
{\bf x})^2}{(\dot {\bf x}^2)^3}\,{.}
\end{equation}

Functional (3.1) is invariant under the following transformations:
\begin{itemize}
\item[i)] translations of the curve coordinates by a constant vector
\begin{equation}
 {\bf x}\rightarrow  {\bf x} +  {\bf a}, \quad  {\bf a} =
 \mbox{const.}
\end{equation}
\item[ii)] $SO(D)$-rotations of the ambient space coordinates
\begin{equation}
 x^i \rightarrow  x^i  +  \omega^{ij} x^j,  \quad  \omega^{ij} = -
 \omega^{ji}, \quad i,j = 1,2,... ,D;
\end{equation}
\item[iii)] reparametrization
\begin{equation}
 \tau \rightarrow f(\tau)
\end{equation}
with an arbitrary function $f(\tau)$ subjected to the condition $\dot
f (\tau) \neq 0$.
\end{itemize}

 According to the first Noether theorem, the
invariance of the functional (3.1) under the translations (3.3)
entails the conservation, under the motion along the curve $ {\bf
x} (s)$, of the `momentum' vector
\begin{equation}
\label{density}
P^i =\frac{d}{d \tau}\,\left (  \frac{\partial \sqrt {\dot {\bf x}
^2}\,\rl (k_1)}{\partial \ddot x^i } \right ) - \frac{\partial
\sqrt {\dot {\bf  x}^2}\,\rl (k_1)}{\partial \dot x^i }\, \equiv
P_{0}^i{,} \quad i = 1,2, \ldots, D,
\end{equation}
where $P_{0}^i$ are constants. The Euler-Lagrange equations (2.3),
rewritten in terms of $P^i$, exhibit this manifestly
\begin{equation}
\frac{d}{d \tau}\,P^i  = 0, \quad i = 1,2, \ldots , D.
\end{equation}

It is clear that the relations (3.6), due to their vector
character, can not be used directly as integrals for Eqs.\
(2.22)--(2.24) determining the principal curvature $k_\alpha$.
Simply, the vector $P^i$ can not be expressed only in terms of the
invariants  $k_\alpha (s)$. However, one can hope that the
invariant
\begin{equation}
{\bf P}^2 = {\bf P}_{0}^2 \equiv M^2,
\end{equation}
 where $M^2$ is a new non negative
constant, can be expressed through the principal curvatures $k_\alpha
(s)$ only. It turns out that this is really the case.

To show this, we pass in Eqs. (3.6) from the differentiation with
respect to arbitrary parameter $\tau$ to the differentiation with
respect to the natural parameter $s$ (the length of a curve). For
this purpose the formulas
$$
 \frac{d}{d\tau } = \sqrt {\dot {\bf
x}^2}\,\frac{d}{ds\,}{,}\qquad \frac{d^2 {\bf x}}{ds^2\,}
  = \frac{\dot{\bf x}^2\ddot {\bf x} -(\dot {\bf x}\, \ddot {\bf x})\,
\dot {\bf x}}{(\dot {\bf x}^2)^2}\,{,} \qquad k_1\,\frac{\partial
k_1}{\partial \ddot {\bf x}} =  \frac{1}{\dot {\bf
x}^2}\,\frac{d^2 {\bf x}}{ds^2\,}{,} $$
\begin{equation}
 k_1\,\frac{\partial k_1}{\partial \dot {\bf x} } =
\frac{1}{(\dot {\bf x}^2)^4}\, \left \{ \left [ 3\,(\dot {\bf x}\,
\ddot {\bf x})^2- 2\, \dot {\bf x}^2\, \ddot {\bf x}^2\right ]
\dot {\bf x}-\dot {\bf x}^2\,(\dot {\bf x}\, \ddot {\bf x})\,\ddot
{\bf x}
 \right \}
\end{equation} are useful. As a result, Eq.\ (3.6) for
the vector ${\bf P}$, constant along the curve ${\bf x}(s)$,
acquires the form
\begin{equation}
{\bf P}  = [2\,\rl '(k_1)\,k_1-\rl (k_1)]\frac{d{\bf x}}{ds\,}+
 \frac{d}{ds\,}\left (
 \frac{\rl '(k_1)\,}{k_1}\right )\frac{d^2 {\bf x}}{ds^2\,}
+\frac{\rl '(k_1)}{k_1}\,\frac{d^3 {\bf x}}{ds^3\,}\,{.}
\end{equation}
The definitions of the parameter $s$ (2.1) and the curvature $k_1$
(2.2) entail the relations
\begin{equation}
\left (\frac{d{\bf x}}{ds\,}\frac{d^2 {\bf x}}{ds^2\,}\right ) =
0, \quad \left (\frac{d{\bf x}}{ds\,}\frac{d^3 {\bf
x}}{ds^3\,}\right ) = - k_1^2,\quad \left (\frac{d^2 {\bf
x}}{ds^2\,}\frac{d^3 {\bf x}}{ds^3\,}\right ) = k_1
\frac{dk_1}{ds\,}{.}
\end{equation}
The vector ${\bf x}'''$ squared can be deduced from the definition
of the second curvature $k_2$ or torsion \cite{Geom,Postnikov}
\begin{equation}
k_1^4\,k_2^2 = {\det}_G\left (\frac{d{\bf x}}{ds\,},\frac{d^2 {\bf
x}}{ds^2\,}, \frac{d^3 {\bf x}}{ds^3\,}\right ),
\end{equation}
where ${\det}_G\left ( {\bf a},{\bf b},{\bf c} \right )$ is the Gramm
determinant for vectors ${\bf a},{\bf b}$, and ${\bf c}$ \cite{KK}.
Thanks to Eq. (2.27), we obtain from (3.12)
\begin{equation}
\left ( \frac{d^3 {\bf x}}{ds^3\,}\right )^2 = k_1^4 +
\left(\frac{dk_1}{ds\,} \right)^2 + k_1^2\,k_2^2 = k_1^4 +
\left(\frac{dk_1}{ds\,}\right)^2 + k_1^2\,\frac{C^2}{(\rl'
(k_1))^4}.
\end{equation}
 Squaring Eq.\ (3.10) and using Eqs.\ (3.11) and (3.13), we have
\begin{equation}
M^2 = \left (\rl'(k_1)\,k_1- \rl (k_1)\right )^2 +
\frac{C^2}{(\rl'(k_1))^2}+(k_1')^2(\rl ''(k_1))^2\,{.}
\end{equation}
In contrast to pseudo-Euclidean space~\cite{NFS1}, the right-hand side of
Eq.\ (3.14) is strictly nonnegative.

By direct differentiation of Eq.\ (3.14) with respect to $s$ one
can be convinced that, for
\begin{equation} \rl''(k_1) \not=  0\,{,}
\end{equation}
the relation (3.14) is an integral of the nonlinear
differential equation (\ref{final}), which determines the
curvature of a stationary curve. From (3.14) we deduce
\begin{equation}
\frac{dk_1}{ds} = \pm\,\sqrt{f(k_1)}\,{,}
\end{equation}
where
\begin{equation}
f(k_1) = \frac{1}{(\rl''(k_1))^2}\left [ M^2 -\frac{C^2}{(\rl
'(k_1))^2}-\left ( k_1\,\rl'(k_1)-\rl(k_1)\right)^2\right ]\,{.}
\end{equation}
Integration of (3.16) gives
\begin{equation}
\int_{k_{10}}^{k_1(s)}\!\!\frac{dk}{\sqrt{f(k)}} = \pm\,(s-s_0)
\end{equation}
with $k_{10} = k_1(s_0)$.

Thus, if the free energy density $\rl (k_1)$ obeys the condition
(3.15), then the curvature $k_1 (s)$ and the torsion $k_2 (s)$ of the
stationary curve ${\bf x} (s)$ are the functions of the parameter $s$
defined by Eqs.\ (3.18) and (2.27). The case when the condition
(3.15) is not satisfied, i.e., when the free energy density $\rl
(k_1)$ is a linear function of the curvature $k_1 (s)$, will be
considered in the next section.

Closing this section we briefly consider the consequences of the
invariance of the free energy under the transformations (3.4) and
(3.5). According  to the first Noether theorem, the rotation
invariance of the functional (2.4) entails the conservation of the
angular momentum tensor
\begin{equation} M^{ij} = \sum_{\sigma=1}^{2}(q_{\sigma}^{i}p_{\sigma}^{j} -
q_{\sigma}^{j}p_{\sigma}^{i}), \quad i<j\,{,}
\end{equation}
where the canonical variables  ${\bf q}_\sigma$ and ${\bf
p}_\sigma,\;\sigma =1,2$ are defined as follows~\cite{NVV} $$ {\bf
q}_{1} = {\bf x}, \quad {\bf q}_{2} = \dot {\bf x}, $$
\begin{equation}
{\bf p}_{1} = {\bf P} = -\,\frac{\partial (\sqrt {\dot {\bf x}
^2}\,\rl)}{\partial \dot {\bf x}}\,-\,\frac{d{\bf p}_2}{d \tau},
\quad {\bf p}_{2}  = -\,\frac{\partial (\sqrt {\dot {\bf x}
^2}\,\rl)}{\partial \ddot {\bf x}}\,{.}
\end{equation}

As in the case of the translation invariance, the conserved angular
momentum tensor (3.19)  does not provide by itself the integrals for
the equations determining the principal curvatures, while the
invariant $S^2$ constructed in terms of this tensor does
\begin{equation} S^2 = \frac{W}{M^2}\,{,}
\end{equation}
where \begin{equation} W = \frac{1}{2}\,M^{ij}M^{ij}{\bf P}^2\,-
\,(M^{ij} P^j)^2,\quad M^2 = {\bf P}^2\,{.}\end{equation} In
Minkowski space-time the analogous invariant is the spin of a
dynamical system \cite{Schweber}. Direct calculation
gives~\cite{NFS1}
\begin{equation} M^2\,S^2 = k_2^2\,(\rl ')^4\,{.}
\end{equation}
Thus the invariant S specifies the integration constant C in Eq.\
(2.27), namely
\begin{equation} C^2 = M^2\,S^2\,{.}
\end{equation}

According to the second Noether theorem \cite{BM}, the invariance of
the functional (3.1) under the reparametrization (3.5) results in the
identity which is obeyed by the left-hand sides of the Euler-Lagrange
equations (2.3). As in the theory of relativistic strings
\cite{BMbook}, one can show that the projection of Eqs. (2.3) on the
tangent vector
 $\dot {\bf x} (\tau)$ is identically equal to zero
\begin{equation} \frac{\delta F}{\delta x^i}\, {\dot  x}^i = 0.
\end{equation}
Thus between equations (2.3) there are only $D-1$ independent ones.
It is this fact that explains why we succeeded in obtaining the
closed set of $D-1$ equation for the principal curvatures of a
stationary curve.

 \section{Determining the functional form of the free energy density}
\setcounter{equation}0

We are going to fix the function $\rl (k_1)$, requiring that all
the solutions to the Euler - Lagrange equations are helices. From
the differential geometry of curves \cite{Geom,Aminov} it is known
that the helices in three dimensional space have a constant
curvature $(k_1)$ and a constant torsion $(k_2)$ which determine
the radius $R$ and the step $d$ of a helix\footnote{It is
interesting to note that for an arbitrary dimension $D$ of the
ambient space, the curves with constant principal curvatures have
a drastically different behavior in the large depending on whether
the dimension $D$ is even or odd. If $D$ is even, then the curves
under consideration are situated in a restricted part of the
space. For odd $D$, such curves go to infinite in one direction.}
\begin{equation}
R = \frac{k_1}{k^{2}_1 + k^{2}_2}, \quad d = \frac{2\pi
|k_2|}{k^{2}_1 + k^{2}_2}.
\end{equation}
As  shown in the previous section, under the condition (3.15) the
curvature $k_1$ and the torsion $k_2$ of the stationary curve are not
constant but they are the functions of the parameter $s$ which are
defined in Eqs.\ (3.18) and (2.27). Hence, for the free energy
density $\rl (k_1)$ we are looking for, we have to replace Eq.\
(3.15) with the requirement
\begin{equation}
\rl'' (k_1) = 0.
\end{equation}
So $\rl (k_1)$ should be a linear function of the curvature $k_1 (s)$
\begin{equation}
\rl (k_1) = \alpha + \beta \, k_1(s),
\end{equation}
where $\alpha$ and $\beta$ are constants. Substituting Eq.\ (4.3)
into (2.27) and (2.28) we obtain
\begin{equation}
k_1 = - \frac{C^2}{\alpha  \beta^3}, \quad  k_2 =
\frac{C}{\beta^2}\,{.}
\end{equation}
Since $k_1 (s) = |{\bf x}'' (s)| > 0$, the constants $\alpha$ and
$\beta$ should have opposite signs. It is natural to put $\alpha > 0$
and $\beta < 0$.

For the free energy density $\rl (k_1)$, linear in curvature, the
integral (3.14) gives just the relation between the integration
constants $M^2$ and $S^2$ (or between $M^2$ and $C^2$)
\begin{equation}
M^2 =  \frac{\alpha^2}{1 -  \beta^{-2} S^2}.
\end{equation}

When $\alpha = 0$, Eq.\ (2.28) implies that the curvature $k_1 (s)$
is an arbitrary function of $s$ and the integration constant $C$
vanishes. In this case Eq.\ (2.27) yields $k_2 = 0$. Hence, for
$\alpha = 0$, the solutions to the Euler-Lagrange equations are
arbitrary plane curves, which is evidently unacceptable for our
purpose.

 Finally, requiring that the stationary curves for the
functional (2.4) are only helices, we uniquely determine  the free
energy density, namely it should be a linear function of the
curvature
\begin{equation}
\rl (k_1) = \alpha - |\beta|\, k_1(s)
\end{equation}
with nonzero constants $\alpha$ and $\beta$, providing $\alpha > 0$
and $\beta < 0$.

In order to get some insight into the physical meaning of the
invariants $M^2$ and $C^2$ we calculate them for protein chains of
different forms. For straight line chain we have $$ M^2 =
\alpha^2, \quad C = 0. $$ For stationary chains (helices) these
invariants are $$ M^2 = \alpha^2\left [1 + \left( \frac{2 \pi
R}{d} \right)^2\right ], \quad C = \alpha\beta \,\frac{2 \pi
R}{d}.
$$

     Here it is worth noting the following.  As known, in the protein
physics the chirality~\cite{Harris} property of these  molecules are of importance.
At the first glance we have ignored this point because from the very
beginning we have eliminated the dependence of the free energy
density $f$ on the curve torsion. But the fact is the integration
constant $C$ in our equations is responsible for the chirality
property of the helical curves under consideration. Really, this
constant is an analog of the Pauli-Lubanski
pseudoscalar\cite{N1,N2} (in the three dimensional space-time
the Pauli-Lubanski vector reduces to the pseudoscalar). The positive
and negative values of this constant distinguish the left-hand and
right-hand chirality of the helical curves.

 Recently, the
functionals defined on the world trajectories in Minkovski
spacetime and depending on the Cartan curvatures of those lines
have been considered to be a generalization of the relativistic
model of a point particle with possible application to string
theory~\cite{rstr}, boson-fermion transmutations \cite{BF}, anyon
models~\cite{anyon}, random walks \cite{random} and so on. It is
worth noting that the Lagrange functions linear in a curvature of
the world line prove to be distinguished in this field too. The
richest set of appealing properties is provided by the Plyushchay
model \cite{Pl,BGPRR}, where $\alpha = 0$. In this case the action
is scale invariant. A complete set of constraints in the phase
space of this model possesses many symmetries, specifically, local
$W_3$ symmetry. This model describes the massless particles, the
helicity of which acquires, upon quantization, integer and
half-odd-integer values. It is interesting that classical
solutions in this model are the spacelike helical
curves~\cite{Pl}.

 \section{Conclusion} \setcounter{equation}0

Proceeding from rather general principles and making use of the
basic facts concerning the conformation of globular proteins we
have obtained, in a unique way, a geometrical model for
phenomenological description of the free energy of helical
proteins. It is worth noting that our functional (4.6) should be
considered as an effective  free energy of the helical protein
which already takes into account the n atomic interactions within
the protein and with the solvent. Hence, there is no need to
quantize it, as one proceeds in the random walk
studies~\cite{random}.

Certainly our simple model does not pretend to describe all the
aspects of the protein folding. However, one can hope that it
could be employed, for example, in Monte Carlo simulation to
search for a stable native state of the protein. In this case the
model can be used for the description of the free energy of
individual parts (blocks) of a protein chain that have the helical
form. Without any doubt, it should result in simplification and
acceleration of the exhausting searching of the native stable
state of the protein chain by a computer~\cite{CD}.

In the general problem of the protein folding, a self containing
task is to reveal the mechanism of the protein chain transition
into the stable native state. A typical time of this process is
such that it is completely insufficient to show the mutual
influence of all the parts of the protein chain during the folding
to the stable state. Presumably the functionals proposed in this
paper can be useful here too. For this purpose the coordinates of
the protein chain ${\bf x}$ should be considered as time
dependent, i.e., ${\bf x} = {\bf x} (t,s)$. During its motion, the
protein chain sweeps out a two dimensional surface in ambient
space described parametrically by its coordinates ${\bf x} (t,s)$.
In this case the analogue of the line curvature $k(s)$ is the
local geometrical invariant of a surface, its external or mean
curvature $H(t,s)$~\cite{Geom}. A straightforward generalization
of the functional (2.4) to the dynamical problem at hand is the
action $S$ linear in external curvature of the surface
\begin{equation}
S = a \iint d\sigma + b \iint H d\sigma,
\end{equation} where $d\sigma$ is a differential
element of the surface swept out by a protein chain in the course
of its motion,  $a$ and $b$ are constants.

In the elementary particle theory the models like (5.1) are known
as the rigid relativistic string \cite{BMbook,KhN}. One of the
motivations to consider such strings was the attempt to develop
string description for quantum chromodynamics. A peculiarity of
the classical dynamics of such string models is its instability
\cite{Instab}. The reason of this is the dependence of the action
(5.1) not only on the velocity $d{\bf x} (t,s)/dt$ but also on the
acceleration $d^2 {\bf x} (t,s)/dt^2$ of the protein chain. As a
result, the energy of such systems proves to be unbounded from
below~\cite{ChN}. It is very likely that this instability could be
crucial in describing, in the framework of the string model (5.1),
the transition of the protein chain to the stable native state.

\section*{Acknowledgements}
This study has been conducted during the stay of one of the
authors (V.V.N.) at Salerno University. He would like to thank G.
Scarpetta, G. Lambiase, and A. Feoli for the hospitality extended
to him. One of the authors (A.F.) wishes to thank A. Rampone for
useful discussions and references about the protein folding
problem. The financial support of INFN is acknowledged. V.V.N.\
was partially supported by the  Russian Foundation for Basic
Research (Grant No.\ 03-01-00025) and by the International Science
and Technology Center (Project No.\ 840).

\end{document}